# Evaluation d'une application de transmission d'images médicales avec un réseau sans fil

*LIIA-120-122, rue Paul Armangot 94400 Vitry Sur Seine-
Université de Créteil (France)*

francomme@univ-paris12.fr, mercier@univ-paris12.fr, ms.chebira@univ12.fr

**Résumé:** Nous proposons une plate-forme de consultation de bases de données et/ou d'échanges d'images biomédicales, adaptée à la transmission bas débit et destinée à des médecins généralistes ou spécialistes. La visée peut-être préventive, diagnostique et thérapeutique. Elle concerne des spécialités telles que la radiologie, l'échographie, l'anatomo-pathologie ou l'endoscopie. Les principales fonctionnalités requises dans un tel contexte sont d'adapter la compression à la fois aux besoins spécifiques de la télémédecine et aux capacités restreintes de communication des réseaux.

Nous présenterons notre démarche dans laquelle nous avons, énoncé des critères sur la qualité des images biomédicales compressées par la méthode des ondelettes permettant de conserver l'ensemble des informations nécessaires à un diagnostic précis, et déterminé les caractéristiques d'un réseau sans fil ayant des performances suffisantes pour la transmission de ces images en respectant les contraintes liées à la modalité et au débit, en l'occurrence la norme 802.11.

Nos résultats permettront d'évaluer la capacité en terme de débit, de cette norme, à transmettre les images à la cadence de 10 images/seconde. Pour cela, il sera nécessaire de quantifier la quantité d'informations à rajouter aux données d'une image afin de permettre une transmission dans de bonnes conditions, ainsi que le mode de fonctionnement approprié.
**Mots clés:** compression, ondelettes, réseau, sans-fil, transmission, Wi-Fi.

## 1 Introduction

Le télédiagnostic nécessite de plus en plus d'échanges d'images médicales de modalités numériques et de volumes (IRM, Scanner X, médecine nucléaire, etc.) entre structures de soins publiques (centres hospitaliers universitaires ou départementaux disposant de haut débit et permettant à plusieurs experts distants d'émettre un avis pour une meilleure prise en charge du patient. La compression des images devient alors une nécessité afin d'assurer leur transfert sur des réseaux de débit limité mais également l'archivage (systèmes connus sous le nom de **PACS** – *Picture Archiving and Communication Systems*) . Le transcodage des images utilisé est alors généralement sans perte, c'est-à-dire permettant une compression réversible. Les taux de compression sont alors réduits par rapport à des techniques dites avec pertes.

Cependant, le transport de ces images jusqu'à la structure de soins qui est le plus souvent directement reliée au patient (cabinet du médecin généraliste ou du spécialiste) implique généralement un transcodage (aujourd'hui principalement au format JPEG) avec une compression irréversible engendrée par le besoin de pouvoir transmettre sur des réseaux à bas débit.

Toutefois la qualité des images ainsi reçues par le médecin est incompatible avec le fait de valider ou d'infirmer un diagnostic, et les images reçues ne pourront pas être gardées dans le dossier patient car seule une version compressée réversible est légalement utile.

De plus, le type de compression étant directement lié à la modalité des images médicales comme cela est indiqué dans le *Tableau 1*, il est nécessaire d'adapter celui-ci au contexte d'exploitation. On ne pourra donc obtenir la même qualité sur toutes les modalités. Dans le cadre de notre travail, nous utilisons des images d'IRM cardiaque et de radiologie thoracique numérique obtenues auprès du laboratoire LISA d'Angers ainsi qu'une base d'images échographiques cardiaques et vasculaires.

Les paramètres permettant de caractériser la qualité des images obtenues dans les différentes modalités sont les suivants :

√ La résolution spatiale → mesure la capacité à distinguer les points d'un objet. Elle correspond au nombre de pixels par unité de surface de l'image.

- √ La résolution de contraste → mesure l'aptitude à distinguer de petites différences d'intensité. Elle correspond au nombre de bits pas pixel.
- √ La résolution temporelle → mesure le temps nécessaire pour créer une image. Une application temps réel exige une génération de 30 images par seconde.

|  | Radio | Médecine nucléaire | Echographie | I.R.M. |
|---|---|---|---|---|
| Résolution spatiale | *Haute* | *Faible* | *Faible* | *Haute* |
| Résolution de contraste | *Faible* | *Bonne* | *Moyenne* | *Haute* |
| Résolution temporelle | *Haute* | *Faible* | *Haute* | *Faible* |

**Tableau 1.** *Caractéristiques de quelques techniques d'imagerie*

On comprend donc l'importance de ces caractéristiques initiales qui ne doivent pas être amputées lors des traitements de compression et décompression, indispensables pour la transmission sur des réseaux bas débit.

## 2 Recherche d'un compromis qualité/ compression

### 2.1 Généralités sur la compression des images

La compression d'images peut s'employer avec des contraintes et des attentes très différentes selon l'utilisation à laquelle les images sont destinées.

On peut vouloir réduire le nombre de bits d'une image avec une contrainte sur la capacité de stockage, la qualité de l'image compressée, la vitesse de transmission, le temps d'accès depuis un médium de stockage, le délai de traitement, la complexité, la compatibilité, le comportement vis-à-vis des erreurs, etc. Son importance est surtout due au décalage qui existe entre les possibilités matérielles des dispositifs que nous utilisons (débits sur Internet, sur Numéris ou sur les différents réseaux, capacité des mémoires de masse,…) et les besoins qu'expriment les utilisateurs pour des applications de loisir, du traitement d'image, de la recherche visuelle rapide dans une base d'images, du diagnostic médical, etc. (visiophonie, vidéo plein écran, TV numérique, stockage et archivage, transfert de quantités d'information toujours plus importantes dans des délais toujours plus brefs).

Les méthodes de compression s'appuient sur deux principes pour atteindre cet objectif, qui sont la représentation approximative de l'information contenue dans l'image (information pertinente introduisant une perte) et la réduction des redondances (n'introduisant pas de perte), au plan spatial et au plan temporel. On peut également considérer la transformation, mais celle-ci ne permet pas de diminuer le nombre des informations, mais de regrouper l'énergie de l'image. Ces principes sont utilisés conjointement dans un schéma complet de codage (*Figure 1*).

Voici une liste synthétique des principales méthodes utilisées :

- √ représentation approximative de l'information ou extraction de l'information pertinente : seuillage, quantification des coefficients, quantification vectorielle, technique de quantification adaptative, sous échantillonnage spatial et temporel.
- √ Réduction de redondances :
  - **Codage prédictif** : modulation Delta, MICD (DPCM) ligne par ligne, MIC 2-D (DPCM 2-D), prédiction inter-trame.
  - **Décorellation** : transformée en cosinus discret (DCT), codage en sous-bandes, codage par les ondelettes, codage par transformation 3-D.
  - **Codage entropique et codage des longueurs de plage** : codage Huffman, codage arithmétique, codage par plan de bits.

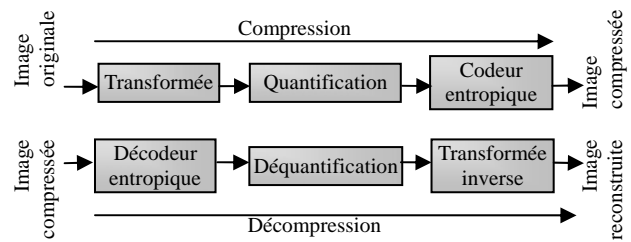

**Figure 1.** *Schéma de principe de la compression par transformation*

### 2.2 La redondance

L'existence de dépendances statistiques entre les échantillons dans une image et entre les images (intra et inter-images) conduit à une redondance de l'information à deux niveaux :

- √ Spatial : corrélation de l'information sur des pixels voisins.
- √ Temporel : corrélation des échantillons des images successives dans une séquence vidéo.

### 2.3 Le manque de pertinence de l'information

Ce manque de pertinence est lié aux propriétés de la perception humaine qui est imparfaite. Par exemple, la perception visuelle humaine possède une résolution limitée des :
- √ Structures proches des fortes transitions de niveau (masquage spatial)
- √ Des objets en mouvement (masquage temporel)
- √ Du mouvement (résolution temporelle)
- √ Des valeurs des pixels (résolution spectrale)
- √ De structures statiques (résolution spatiale)

Ainsi une méthode de compression tend à extraire l'information pertinente et à supprimer les redondances statistiques.

Les tableaux 2 et 3 présentent respectivement des tailles de fichiers numériques pour des images fixes et des séquences vidéo, de différentes résolutions.

La problématique de la compression d'image consiste donc à satisfaire l'ensemble des contraintes tout en obtenant la qualité requise de l'image décompressée pour l'application désirée.

|  | Résolution (l*L) | nbpp | Taille fichier |
|---|---|---|---|
| Image en niveau de gris | 512 * 512 | 8 bits | 2Mbits ou 256Ko |
| Image en couleur | 512 * 512 | 8*3 bits | 6 Mbits ou 768 Ko |

**Tableau 2.** *Taille de fichiers numériques pour des images fixes*

|  | Résolution | nbpp | Débits bruts Pour 25 images/s | Débits utilisés en pratique |
|---|---|---|---|---|
| Real video | 160*120 | 12 bits | 0.2 Mbits/image 5.5 Mbits/s ou 0.7 Mo/s | 0.2 Mbits/s |
| DVD | 720*576 | 12 bits | 4.8 Mbits/image 120 Mbits/s ou 15 Mo/s | Moy. 4.5 Mbits/s Max. 9.8 Mbits/s |

**Tableau 3.** *Taille de fichiers numériques pour des séquences vidéo*

### 2.4 Mesures des performances

**Taux de compression**

Une mesure courante pour déterminer le degré de compression obtenu est le taux de compression *CR*1. Il est défini par :

$$CR = \frac{\text{nombre de bits de l'image originale}}{\text{nombre de bits de l'image comprimée}} = \frac{R_0}{R_c}$$

Pour une même méthode de compression et un même *CR* réalisés sur des images distinctes, la qualité obtenue peut être très variable d'une image à l'autre. Les propriétés statistiques des images originales jouent un rôle prépondérant dans le résultat obtenu.

Par exemple avec une image sur-échantillonnée, donc très redondante, il est facile d'obtenir des taux élevés. La théorie de l'information donne une limite théorique au *CR* maximal qu'il est possible d'obtenir sans distorsion pour toute méthode de compression sur une image donnée (Shannon, 1948).

**Entropie**

L'entropie est une grandeur qui caractérise la quantité d'informations que contient une image. Par exemple une image dont tous les pixels ont la même valeur contient très peu d'informations car elle est extrêmement redondante, son entropie est faible. En revanche, une image dont tous les pixels ont une valeur aléatoire contient beaucoup d'informations ; son entropie est forte.

En pratique, l'entropie d'une image numérique est inversement liée à la probabilité d'apparition des niveaux de gris dans l'image. Plus une valeur de gris k

---

1 **CR** : de l'anglais "compression ratio" (taux de compression)

---

est rare, plus sa probabilité d'apparition *p(k)* est faible, et cela contribue à une entropie globale plus grande. Par définition, l'entropie d'ordre zéro *H0* est donnée par :

$$H_0 = -\sum_{k=0}^{2^R-1} p(k).\log_2 p(k) \qquad \text{unité } bpp$$

L'utilisation du logarithme de base deux, fait de *H0* le nombre moyen de bits par pixel nécessaire pour coder toute l'information contenue dans l'image. Une image codée avec *R* bits par pixels a en fait presque toujours une entropie d'ordre zéro *H0* inférieure à *R*.

Par conséquent, l'entropie *H0* d'une image originale fournit le débit minimal qu'il est possible d'atteindre par compression pixel par pixel sans dégrader l'image, et par là même un taux de compression sans perte maximal (*Figure 2*).

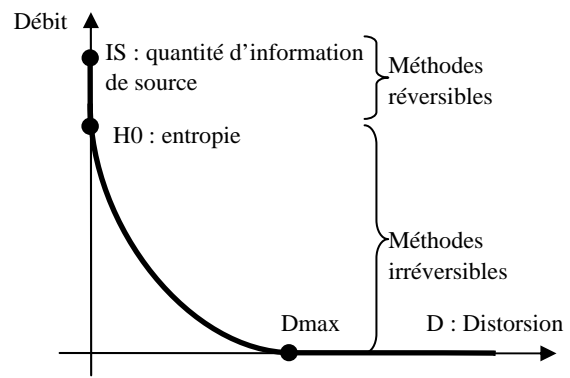

**Figure 2.** *Fonction débit/distorsion*

La fonction débit/distorsion présente deux zones distinctes. La première est la zone sans distorsion (*CR*<2.5). Dans cette zone la courbe est confondue avec l'axe des ordonnées puisque *D=0*, la restitution de l'image originale est possible. La seconde est la zone avec distorsion (*CR*>2.5). Les images sont ici codées avec un nombre de digits binaires inférieur à l'entropie *H0*. Une distorsion est alors introduite dans l'image décodée qui ne sera pas identique à l'image originale.

### 2.5 Mesures de la distorsion

*Mesures objectives de distorsions dues à la quantification*

La distorsion *D* est l'erreur introduite par l'opération de compression, due au fait qu'éventuellement l'image reconstruite n'est pas exactement identique à l'image originale.
La mesure de distorsion utilisée généralement en compression d'image est l'*erreur quadratique moyenne*2. Cette grandeur est définie par la moyenne des écarts au carré $e_{mn}^2$ entre le pixel (m,n) de l'image originale I(m,n), et le pixel (m,n) de l'image reconstruite Î(m,n). M*N le nombre total de pixels de l'image.

$$MSE = \frac{1}{M.N} \sum_{m=0}^{M-1} \sum_{n=0}^{N-1} \left[ I(m,n) - \hat{I}(m,n) \right]^2$$

On référence cette erreur par rapport à la dynamique de l'image en décibels. On obtient un rapport signal sur bruit crête pour une image dont le maximum est dénoté PSNR (peak SNR3). Si le minimum de l'image est nul (image bien cadrée) on obtient le rapport signal sur bruit crête à crête noté PP SNR (peak to peak SNR).

Lorsque la reconstruction est parfaite, la MSE est nulle et le PPSNR tend vers l'infini.

$$PSNR = 10.\log_{10}\frac{(2^R-1)^2}{MSE} \text{ dB}$$

Ces mesures de distorsion sont objectives et simples à calculer. Certaines méthodes de compression recherchent le meilleur compromis entre la performance et la distorsion, et optimisent des courbes taux-distorsion ou *R(D)4*.

L'inconvénient de la MSE est qu'elle ne rend pas compte de la perte de qualité visuelle engendrée par la compression. Si tous les pixels d'une image étaient translatés, l'erreur quadratique serait très élevée, alors que la qualité visuelle serait parfaitement bonne. De plus, la MSE est une mesure globale sur toute l'image, qui gomme les variations locales.

Par exemple, dans une image médicale, si des détails anatomiques importants sont dégradés par la compression et si la majeure partie du reste de l'image est fidèlement restituée, alors la MSE est relativement faible, mais pour l'expert médical, cette image a une qualité diagnostique médiocre. De très nombreuses recherches visent à trouver des mesures objectives de distorsion qui prédisent suffisamment bien la qualité perceptuelle. Ces travaux ont apporté une connaissance sur les réponses du système visuel humain (dénoté *HVS*5) à certaines formes de dégradation. Il est de plus nécessaire de valider la mesure de distorsion à l'utilisation et l'usage qui sera fait des images décompressées. Ces études, même appliquées à l'imagerie médicale, n'ont pas encore abouti à une mesure de distorsion qui prédise de façon satisfaisante la qualité des images reconstruites en fonction de leur emploi pour une interprétation visuelle ou automatique, avec ou sans analyse quantitative (par exemple dans l'évaluation des sténoses). L'état de l'art consiste encore à se baser sur des mesures de distorsion lors du développement de la méthode de compression, et ensuite de le valider à l'aide d'observateurs par des comparaisons statistiques. Cette approche largement utilisée dans notre travail est basée sur des résultats obtenus par le LISA CNRS-FRE 2656 Groupe signal et image – Centre Hospitalier Universitaire d'Angers et présentés dans la publication « *Mise en place de paramètres quantitatifs caractérisant la dégradation engendrée par une chaîne de compression en ondelettes (JPEG2000)* ». Source : Christine Cavaro-Menard, maître de conférence (Cavaro_Menard, 2004).

*Mesures subjectives de distorsions*

Elles visent à définir la perception de la dégradation d'une image. Elles nécessitent :

√ une expérimentation psycho-visuelle

√ un modèle fin du système de vision humain

## 3 Les réseaux sans fil dans les applications médicales

### 3.1 Les avantages des réseaux sans fil

Aujourd'hui, le monde de l'entreprise en général se caractérise par un fort développement de la « nomadicité ». Les employés sont équipés d'ordinateurs portables et passent plus de temps à travailler au sein d'équipes plurifonctionnelles et géographiquement dispersées.

La productivité des employés est pour une grande part générée au cours de réunions, et non pas sur le poste de travail. L'utilisateur doit pouvoir accéder au réseau ailleurs qu'à son poste et le réseau local sans-fil (Wireless LAN ou WLAN) s'intègre parfaitement dans cet environnement, offrant aux utilisateurs mobiles la liberté d'accès au réseau dont ils ont besoin.

Quelle que soit sa taille, l'entreprise peut bénéficier du déploiement d'un système WLAN, qui associe le débit des réseaux filaires, l'accès mobile et la souplesse de configuration.

De façon générale, les réseaux sans-fil permettent un gain en termes :
√ de réduction des coûts sur les alternatives filaires, une installation plus économique du réseau dans les endroits difficiles à câbler : bâtiments anciens et structures en béton armé mais plus particulièrement dans les environnements dynamiques nécessitant des transformations.
√ de réactivité, d'efficacité organisationnelle, de productivité : une mobilité génératrice de gains de productivité, avec un accès en temps réel aux informations, quel que soit le lieu où se situe l'utilisateur (salle de conférence, cafétéria ou site distant), pour une prise de décision plus rapide et plus efficace.

Les WLAN libèrent l'utilisateur de sa dépendance à l'égard des accès câblés ou backbone (*Figure 3*), lui offrant un accès permanent et omniprésent. Cette liberté de mouvement offre de nombreux avantages dans de nombreux types d'environnements de travail tels que :

√ accès immédiat entre le lit d'hôpital et les informations concernant le patient pour les médecins généralistes et/ou spécialistes et le personnel hospitalier.

√ Un accès réseau simple et en temps réel pour les consultants et les auditeurs sur site : réunion de groupes d'étude et liens de recherche pour les étudiants.

---

2 **MSE** : de l'anglais "*Mean Square Error*" (*erreur quadratique moyenne* : **EQM**)
3 **SNR** : de l'anglais "Signal to Noise Ratio"
4 **R (D)** : de l'anglais "Rate-Distorsion"
5 **HVS** : de l'anglais "Human Visual System"

√ Un accès étendu aux bases de données pour les chefs de service nomades, directeurs de chaîne de fabrication, contrôleurs de gestion ou ingénieurs du bâtiment.

√ Une configuration simplifiée du réseau avec un recours minime au personnel informatique pour les installations temporaires telles que stands de foire, d'exposition ou salles de conférences.

√ Un accès plus rapide aux informations client pour les fournisseurs de services et détaillants, résultant en un meilleur service et une satisfaction supérieure.

√ Un accès omniprésent au réseau pour les administrateurs, pour la conception, le déploiement, l'amélioration du réseau et le dépannage.

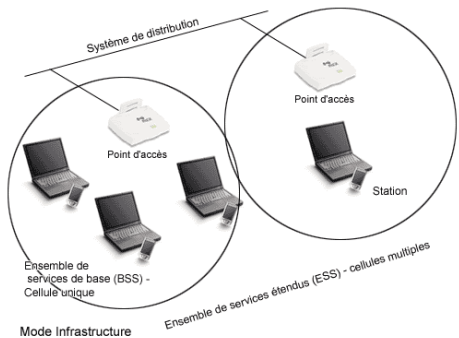

**Figure 3.** *Ensemble de services pour le Wi-Fi*

### 3.2 Les familles des réseaux sans fil

Dans le cadre de notre application, on peut distinguer deux types principaux de réseaux sans fil (*Figure 4*) :

√ Le réseau personnel sans-fil (Wireless Personal Area Network, WPAN) constitué des connexions entre des appareils distants de quelques mètres (PC, assistants, périphériques divers...). Bluetooth dont le rayon d'action ne peut dépasser les 50 mètres est donc strictement destiné au WPAN.

√ Le réseau local sans-fil (Wireless Local Area Network, WLAN) qui correspond donc au périmètre d'un réseau local, installé dans une entreprise, dans un foyer ou encore dans un espace public. Tous les terminaux (PC, assistants) situés dans la zone de couverture du WLAN peuvent s'y connecter. Plusieurs WLAN peuvent s'adosser à une même architecture (le même annuaire, les mêmes paramètres de connexion) de manière à faciliter la connexion d'un utilisateur qui traverse plusieurs zones de couverture. Wi-Fi, HiperLan et HomeRF peuvent émettre jusqu'à 100 mètres, ce qui rend leur usage pertinent pour des réseaux locaux

**Remarque** : Zigbee est la dernière ratification (août 2003) de standard 802.1.5-4 dans la bande ISM – 2.4 Ghz. Son débit maximum est de 250 Kbits/s, mais plus généralement il est limité à 20 Kbits/s en Europe ; par contre la qualité du signal est excellente. Il est surtout conçu pour les réseaux de capteurs orientés faible consommation. Sa portée peut atteindre 100 m mais son débit l'écarte du cadre de notre application.

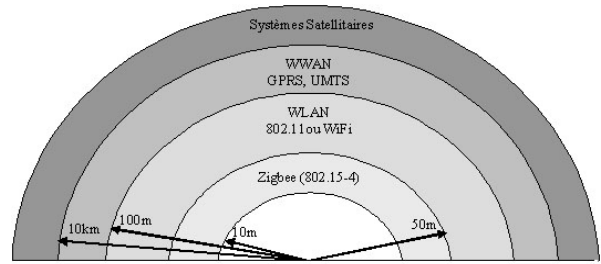

**Figure 4.** *Rayon d'action des réseaux sans fil*

**Remarque** : Les réseaux WWAN et WMAN sortant du cadre de cette étude ne sont pas décrits dans cet article.

### 3.3 Complémentarité des réseaux sans fil

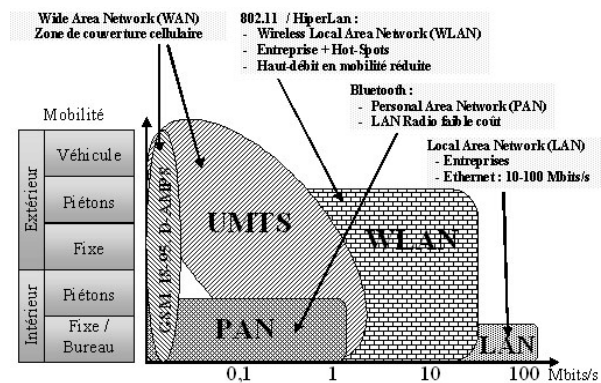

**Figure 5.** *Complémentarité des réseaux sans fil*

### 3.4 Les inconvénients des réseaux sans fil

Tout comme les autres technologies sans fil, les WLAN doivent être conformes aux standards stricts définis par les gouvernements et l'industrie. On s'est beaucoup inquiété dans les industries qui font appel aux technologies sans fil des risques du « sans fil » pour la santé. A l'heure actuelle, aucune étude scientifique n'a permis de déterminer que les transmissions sans fil avaient un effet nuisible sur la santé. De plus, la puissance d'émission des LAN sans fil est limitée à moins de 100 mW par les réglementations de la FCC (*Tableau 4*) ; ce qui est nettement inférieur à la puissance des téléphones mobiles, alors qu'il est généralement admis que les risques causés par les transmissions radio sont corrélés à la puissance et à la proximité de l'émetteur.

### 3.5 Le coût

Le coût matériel doit tenir compte de l'intégration des points d'accès à l'infrastructure du réseau et des cartes WLAN à tous les périphériques et ordinateurs sans fil. Le nombre de points d'accès dépend de la zone de couverture, du nombre d'utilisateurs et des types de services requis. La zone de couverture de chaque point d'accès s'étend à partir de celui-ci sur un rayon donné. Fréquemment, ces "zones" se recouvrent de manière à assurer une couverture complète. Le coût matériel dépend donc de facteurs tels que les performances et la couverture requises, ainsi que des débits.

En plus du coût d'équipement, il faudra aussi prendre en compte les frais d'installation et de maintenance.

### 3.6 Règlementation pour l'utilisation du spectre radio

L'utilisation de ce spectre est très réglementée, notamment parce qu'il est encore exploité par l'Armée. Aux Etats-Unis, la *Federal Communication Commission* (*FCC*) est une agence gouvernementale indépendante qui établit les règles d'utilisation des fréquences et des puissances d'émissions. En Europe, il existe un organisme équivalent à la *Conférence Européenne des Postes et Télécommunications* (*CEPT*), qui dispose de prérogatives équivalentes, à laquelle adhère l'*Agence française de Régulation des Télécommunications* (*ART*). Elle édicte les règlements et en contrôle ses applications sur le territoire français métropolitain et outre-mer. A quelques nuances près, les bandes de fréquences définies aux Etats-Unis sont applicables en Europe. Des négociations sont en cours avec le ministère de la Défense et l'ART (Autorité de Régulation des Télécoms : http://www.art-telecom.fr/) qui prévoient une libéralisation de l'usage. L'ouverture des réseaux radio-électriques (*RLAN*) a toutefois commencé : en juillet 2001, l'ART a précisé les conditions d'utilisation de Bluetooth et d'HiperLan. Ces conditions concernent la puissance des émetteurs et le lieu d'émission. Pour résumer, les RLAN à l'extérieur des bâtiments et sur le domaine public sont interdits ; ils sont en revanche autorisés à l'intérieur des bâtiments et à l'extérieur tant qu'il s'agit d'un domaine privé et tant que les émetteurs respectent des limites de puissance.

La généralisation des WLAN dépend de la standardisation de l'industrie. Celle-ci assurera la fiabilité et la compatibilité des produits entre les divers équipementiers. L'IEEE (*Institute of Electrical and Electronics Engineers*) a ratifié la spécification 802.11, norme régissant les réseaux locaux sans fil, en 1997. Le standard 802.11 s'appliquait à des débits de 1 et 2 Mbps et définissait les règles fondamentales de la signalisation et des services sans fil.

Le principal problème, qui limitait les perspectives de développement de l'industrie du WLAN, était alors ce débit limité, trop faible pour répondre réellement aux besoins des entreprises. Conscient de la nécessité d'augmenter ce débit, l'IEEE a ratifié la spécification 802.11HR (également baptisé 802.11 Haut Débit) qui entérine des transmissions à 11 Mbps maximum. En dehors des organismes de normalisation, les principaux acteurs de l'industrie du sans fil se sont réunis au sein de la WECA (*Wireless Ethernet Compatibility Alliance*). La mission de la WECA est de certifier l'interopérabilité et la compatibilité inter-fournisseurs des équipements pour réseaux sans fil IEEE 802.11HR, ainsi que de promouvoir ce standard auprès des Grands Comptes, des PME et du grand public. La WECA regroupe des fabricants de semiconducteurs pour WLAN, des fournisseurs de WLAN, des fabricants d'ordinateurs et des éditeurs de logiciels. On retiendra entre autres 3Com, Aironet, Apple, Cabletron, Compaq, Dell, Fujitsu, IBM, Intersil, Lucent Technologies, Nokia, Samsung, Symbol Technologies. Avec 802.11HR, les WLAN offriront des performances et un débit comparables à l'Ethernet filaire.

### 3.7 Les fréquences du réseau WiFi

Le *Tableau 4* récapitule les bandes de fréquence sans licence utilisées dans les RLAN aux Etats-Unis et avec restrictions en Europe.

- √ Bande ISM sans licence (Instrumentation Scientific & Medical) de 2.4 Ghz à 2.4835 GHz.
- √ Bande U-NII sans licence de 5.15 GHz à 5.825 GHz environ, avec certaines restrictions d'utilisations en Europe.

|  | Fréquences (GHz) | Puissance en France (mW) | Puissance aux USA (mW) | Bande |
|---|---|---|---|---|
| Utilisation intérieure | 2.4–2.4465 | < 10 | <1000 | ISM |
|  | 2.4465–2.4835 | < 100 | <1000 |  |
|  | 5.15–5.25 | < 50 | <1000 | U-NII |
|  | 5.25–5.35 | < 200 avec conditions | <1000 |  |
| Utilisation extérieure | 2.4–2.4465 | Interdit | <1000 | ISM |
|  | 2.4465–2.4835 | < 100 + Autorisation | <1000 |  |
|  | 5.725–5.825 | Interdit | <1000 | U-NII |

**Tableau 4.** *Bandes de fréquence et puissance des RLAN*

**Les standards du sans-fil**

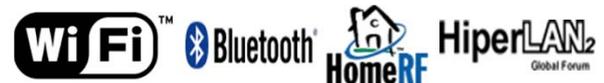

**Figure 6.** *Les technologies de réseaux sans fil*

### 3.8 Avenir du WIFI

A cette date, certaines de ces technologies souffrent de handicaps qui laissent planer des doutes sur leur avenir.

HomeRF, sévèrement concurrencé par Wi-Fi aux Etats-Unis, a perdu le soutien de deux sponsors de poids, à savoir Intel et Microsoft. Du coup, cette norme est vraiment en perte de vitesse.

Bluetooth ne répond pas au problème car il est orienté picoréseaux. De plus, les performances (le débit notamment) semblent faibles au regard des applications de transfert d'images. Une prochaine version prévoit toutefois de multiplier par 10 ces débits. Cette technologie tarde donc à se concrétiser mais profite encore de deux avantages :

- √ elle n'a pas vraiment de concurrent
- √ en France, l'ART a autorisé son exploitation.

Wi-Fi et HiperLan se font face et présentent des évolutions (802.11g et Hiperlan2) directement concurrentes. Le premier est activement soutenu par l'essentiel de l'industrie de la High-Tech américaine (Microsoft permet de l'exploiter depuis Windows XP) tandis que le second bénéficie de soutiens en Europe et, surtout, a reçu l'aval de l'ART en France.
HiperLan a aujourd'hui un rôle marginal dans les technologies radio servant de base à des réseaux locaux. La principale explication de cet échec réside dans l'origine strictement européenne de ce protocole et la complexité de réalisation des circuits qui le rend coûteux et non compétitif par rapport au 802.11.

Les recherches effectuées en ce moment autour de la norme 802.11 tournent autour de 3 axes principaux : haut débit, sécurité, QoS. L'idéal serait de définir un standard regroupant ces 3 aspects en même temps.

***Haut débit*** : 802.11g utilise une technique radio appelée OFDM. Au lieu d'envoyer les bits de données de façon séquentielle à un haut débit, OFDM envoie de multiples flots de données en parallèle. 802.11g opère dans la bande ISM des 2.4 GHz.
La norme 802.11a dénommée Wi-Fi 5 par la WECA, utilise la bande U-NII des 5 GHz qui offre une largeur de bande plus importante (300MHz) et qui est beaucoup moins encombrée que la bande ISM des 2.4 GHz. Par contre, elle est totalement incompatible avec les autres normes physiques.

***Sécurité*** : La méthode de sécurisation de la norme 802.11 appelée WEP (Wired Equivalent Privacy) utilise un codage sur 40 ou 128 bits et utilise l'algorithme RC4. Mais WEP possède de nombreux points faibles (Schwenck, 2002). L'IEEE développe donc actuellement une nouvelle norme de sécurité, 802.1x qui pourra être appliqué à tous les réseaux IEEE (filaire ou non). Cette architecture devrait fournir des systèmes d'authentification, de cryptage, d'intégrité de message et de distribution de clés.
Un autre standard, 802.11i, définit les différents niveaux de sécurité à implémenter dans les WLAN 802.11a, g et b. Ce standard a été ratifié en mars 2004.

***QoS*** : La Qualité de Service (QoS) est définie par le protocole 802.11e. Celui-ci fournit de la QoS pour tous les types de flux de données (audio ou vidéo). Il permet à chaque flux de données d'employer une politique différente. Par exemple, un flux vidéo pourra subir un traitement différent de celui subit par un flux de données classique. La QoS est essentielle pour supporter des flux vidéo ou audio.

## 4 Résultats et discussion

### 4.1 Choix de la méthode de compression

Il est admis que les méthodes de compression par transformation sont les plus efficaces en terme de taux de compression. Parmi celles-ci, la méthode JPEG largement utilisée dans les applications existantes montre ses limites pour les forts taux de compression avec l'apparition de l'effet de bloc (importante dégradation de l'image originale). La méthode de transformation par les ondelettes quant à elle est une méthode globale qui travaille sur l'ensemble de l'image sans découpage et fait apparaître un flou pour les forts taux de compression ; c'est en l'occurrence la technique utilisée dans le format Jpeg2000.

Pour la mise en œuvre de l'application, nous avons choisi d'implémenter l'analyse multi-résolutions par la méthode du codage en sous-bandes. Notre méthode sera donc basée en partie sur les techniques utilisées dans la norme JPEG2000. Nos images sont au format PGM équivalent au format DICOM ou BMP et nous ne considèrerons qu'un codage en 256 niveaux de gris (8 bits).
Nous associerons cette technique à la compression RLE suivie de celle de Huffman.

Pour la compression d'une séquence d'images, le principe adopté s'appuie sur un codage utilisant les relations inter-images. Afin de pouvoir distinguer les différences entre ces résolutions d'images et ratios de compression, nous utilisons dans un premier temps des images non médicales pour lesquelles nous sommes plus aptes à juger de la qualité restituée (i.e. Lena, boat, goldhill,…). L'IRM a été également retenu pour cette étude. En effet la qualité des images obtenues par cette modalité en terme de contraste, de bruit et de finesse de la texture, se prête bien à l'évaluation des performances de la méthode de compression. Les séquences cérébrales sont les plus utilisées et ont donc été naturellement exploitées.

### 4.2 Choix du taux de compression

Dans le contexte particulier de l'imagerie médicale, la compression d'information ne doit pas conduire à des erreurs d'interprétation et de diagnostic. L'évaluation de la qualité de l'image reconstruite ne peut plus alors se contenter des critères classiques : erreur quadratique moyenne, rapport signal sur bruit, écart type, etc.

Un protocole particulier d'évaluation psycho-visuel a été mis en place par le LISA (Cavaro-Menard, 2004). Il a nécessité la collaboration de plusieurs médecins à qui ont été proposées plusieurs séries de séquences compressées à différents taux. De nombreux critères tels que la netteté des sillons et des bords, la distinction entre les surfaces grises et les surfaces blanches, la visualisation des petits vaisseaux, l'apparition de l'effet de flous/blocs ont été soumis à l'appréciation des examinateurs permettant ainsi d'évaluer la qualité visuelle des images reconstruites.

D'après ces études menées sur l'évaluation qualitative de structures anatomiques et pathologiques, il semblerait qu'avec la technique des ondelettes (JPEG2000), le taux de compression n'ayant pas permis la modification d'un diagnostic par les médecins est de 20 (Cavaro-Menard, 2004).

Nous avons donc repris ces résultats qui répondent parfaitement à l'objectif de cette première étape et nous retenons donc **la valeur 20 comme taux de compression acceptable,** dans le cas de la compression des images biomédicales **par la méthode des ondelettes**.

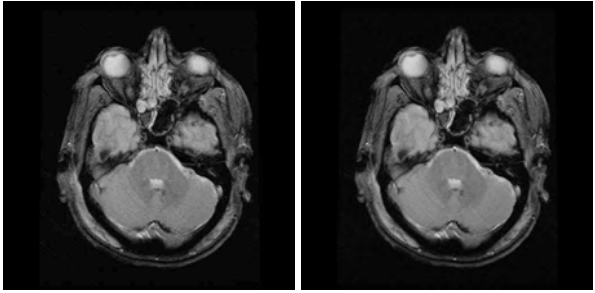

**Figure 7.** *Images compressées respectivement au format JPEG et JPEG2000 avec un CR de 55.*

### 4.3 Validation de la qualité des images médicales

La validation de la qualité des images sera faite dans un premier temps à l'aide des critères mathématiques courants : EQM (*Erreur quadratique moyenne*) et PSNR (*Peak signal/noise Ratio*). Ils seront éventuellement complétés par une appréciation visuelle avec l'assistance de praticiens confirmés.

### 4.4 Taille des fichiers images compressées pour différentes modalités.

Le *Tableau 5* donne la taille des fichiers pour les images biomédicales compressées respectant le critère sur le taux de compression justifié précédemment.

| Modalités | Format | Nbr bits/pix. | CR | Taille compressée (Kbit/Ko) | Nbr Bits/pix. |
|---|---|---|---|---|---|
| IRM | 256*256 | 16 | 20 | 51,2/6,4 | 0,8 |
| IRM | 512*512 | 16 | 20 | 204,8/25,6 | 0,8 |
| Radiologie numérique | 2000*2000 | 16 | 20 | 3125/390,7 | 0,8 |

**Tableau 5.** *Tailles des fichiers compressés pour différentes modalités*

### 4.5 Débit estimé pour la transmission

Le débit estimé nécessaire à la transmission de ces images compressées sur un réseau pour une séquence de dix images/seconde :

**Images de 256*256 : 51,2*10 = 512 Kbit/s**

**Images de 512*512 : 204,8*10 = 2 Mbit/s**

**Images de 2000*2000 : 3125*10 ≈ 30,5 Mbit/s**

### 4.6 Les contraintes du réseau sans fil

La description des différents réseaux sans fil ainsi que les éléments du *Tableau 5* confirme bien que la norme 802.11 est la plus adaptée dans le contexte qui est celui de la transmission d'images biomédicales au niveau LAN, par rapport à ses principaux concurrents que sont HiperLan et Bluetooth.

Ce choix tient compte des aspects liés à la transmission imposée par le canal : les débits associés à la qualité minimum recommandée pour une exploitation d'images médicales, les modes transfert, les modes d'accès, les statistiques d'erreurs, le délai, la bande passante ainsi que le prix de revient d'une installation pour un type de réseau sans fil.

Le type de modulation permet d'obtenir un débit plus ou moins important : 54 Mbit/s théorique maximum pour la norme 802.11g. Chacune de ces modulations est plus ou moins sensible au SNR, ce qui provoque une ajustement automatique du débit lorsque le BER[6] dépasse certains seuils, et inversement lorsque le BER augmente. Ceci est notamment dû au fait de ne plus pouvoir discerner les différents symboles.

|  | 802.11 | 802.11a | 802.11b | 802.11g |
|---|---|---|---|---|
| Standards approuvés | Juillet 1997 | Septembre 1999 | Septembre 1999 | Juin 2003 |
| Bande passante disponible | 83,5 Mhz | 300 Mhz | 83,5 Mhz | 83,5 Mhz |
| Nbr canaux non recouvrant | 3 | 4 | 3 | 3 |
| Débit par canal (Mbps) | 1, 2 | 6,9,12,18,24, 36,48,54 | 1, 2,5.5,11 | 1,2,5.5,9,11,22, 24,33,36,54 |
| Débit par type de modulation (Mbps) | DQPSK (DSSS:2) DBPSK (DSSS:1) 4GFSK (FHSS:2) 2GFSK (FHSS:1) | BPSK (6,9) QPSK (12,18) 16-QAM (24,36) 64-QAM (48,54) | DQPSK/CCK (5.5, 11) DQPSK (2) DBPSK (1) | OFDM/CCK (6,9,12,18,24, 36,48,54) DQPSK/CCK (5.5,11,22,33) DQPSK (2) DBPSK (1) |
| compatible | 802.11 | Wi-Fi5 | Wi-Fi | Wi-Fi |

**Figure 8.** *Les techniques de modulation pour les standards de la norme 802.11*

### 4.7 Choix des hypothèses sur le débit du réseau

La norme choisie pour la mise en œuvre de cette application est le 802.11g capable de fournir un débit théorique maximum de 54 Mbps. Afin de nous placer dans des conditions de fonctionnement reflétant un contexte réaliste et ainsi tenir compte des inconvénients inhérents aux transmissions sans fil (affaiblissement du signal, erreurs de transmission), nous décidons de:

√ considérer un débit moyen de 11 Mbits/s (le cas du 802.11b) étant donné que les conditions permettant la transmission avec le débit de 54 Mbits/s ne sont, sur le terrain, que très peu probable.

√ considérer la retransmission systématique de chaque trame (paquet) de données due a l'altération des trames transmises (Fainberg, 2001).

### 4.8 Choix de la norme et du protocole de haut niveau

Le choix du protocole TFTP se justifie dans sa simplicité de mise en œuvre mais surtout par sa faible surcharge : en effet ce protocole utilisant le protocole simple UDP, la trame envoyée à la couche MAC, est ainsi plus petite qu'avec le protocole TCP. L'inconvénient majeur de ce choix est l'impossibilité de vérifier l'intégrité des données transmises entre la couche transport et la couche MAC car aucune vérification n'est effectuée lors de ce transfert. Ce point sera éventuellement à voir lors d'un travail approfondi sur cette application.

---

6 **BER** : Bit Error Rate

Dans la mesure où nous travaillons dans un environnement relativement peu perturbé pour ce prototype logiciel, aucune perte sensible n'altère les résultats obtenus sur la qualité des images biomédicales.

Cette manipulation pose le problème de la sécurité des informations transitant sur les réseaux sans fil : bien que la norme 802.11g mette en œuvre des mécanismes de sécurité, ceux-ci ne seront pas mis en œuvre dans le cadre de ce travail. On peut toutefois pressentir une baisse des performances dues notamment au temps de calcul non négligeable des algorithmes de cryptographie.

### 4.9 Estimation sur le débit réel

L'objectif ici, est d'estimer le débit qu'il est possible d'obtenir en tenant compte de l'ensemble des contraintes liées aux standards 802.11 ainsi qu'aux imperfections du support de transmission. Celui-ci permettra notamment de régler de façon optimale le temps inter-balise du mode infrastructure pour le standard 802.11 (*Figure 9*).

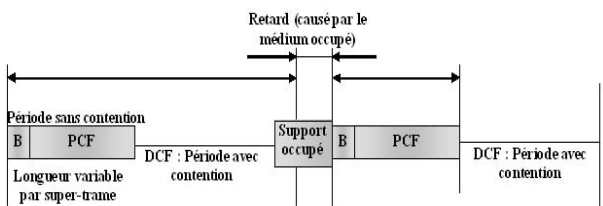

**Figure 9.** *Gestion du temps par le point d'accès*

La valeur par défaut de ce paramètre est 100ms. On prévoit le réglage de celui-ci afin de permettre la réception de l'ensemble des paquets de données d'une même image dans une seule période sans contention (PCF).

Connaissant le débit, il faut ensuite estimer la taille de la trame devant être transmise ainsi que les différents éléments de protocole devant être pris en compte.

La taille après compression pour des images de résolution 512*512, est de 204,8 Kbit ou 25,6 Koctet (*Tableau 5*). Le Protocole TFTP fragmente le fichier de l'image par paquet de 512 octets.

Cette taille de fragmentation est choisie en respectant les spécifications pour ce protocole mais également parce que la transmission par petits paquets de données est recommandée pour les liaisons hertziennes dans la mesure où la perte d'un paquet de données nécessite la retransmission de celui-ci. Une étude réalisée par D. Dhoutaut et I. Guérin-Lassous montre que la fragmentation est également un paramètre important agissant directement sur le débit (Dhoutaut & al., 2003). Je proposerai une estimation du débit en fonction de trois niveaux de fragmentations distincts : 512, 1024, 2048, les deux premiers étant inférieurs au seuil de fragmentation annoncé dans l'article cité précédemment.

### Scénario 1 : le mode Ad-Hoc sans réservation de support

On choisit pour commencer le mode Ad-Hoc (DCF) sans mécanisme de réservation (RTS/CTS). La *Figure 10* illustre le scénario pour lequel on prévoit de transmettre chaque trame de données deux fois et de simuler ainsi des défauts de transmission. C'est le mode fragmentation « burst ».

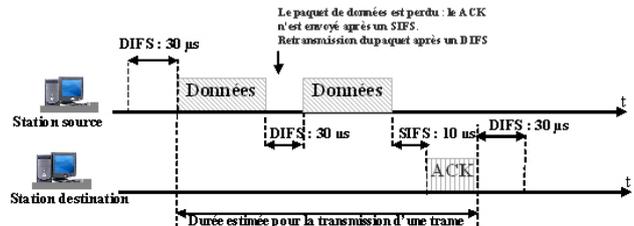

**Figure 10.** *Séquence de transmission en mode DCF avec perte d'une trame*

On obtient par le calcul, pour ce mode, un taux de transfert effectif de **3.16 Mbits/s** (*Figure 14*) pour le fichier image de 25.6 Koctets.

### Scénario 2 : le mode Ad'Hoc avec réservation de support

De la même manière que dans le scénario 1, on peut quantifier la performance en terme de débit pour le mode Ad'Hoc avec un mécanisme de réservation de médium (RTS/CTS). On remarque que l'on trouve deux trames de plus dans la séquence. On vérifie l'impact de celles-ci sur le débit du réseau.

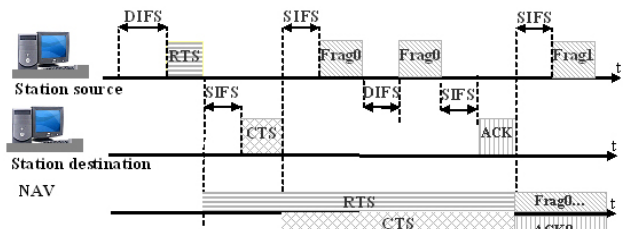

**Figure 11.** *Séquence de transmission en mode DCF avec RTS/CTS*

### Scénario 3 : le mode PCF

Le temps dans la période sans contention étant primordial, les accusés de réception, le polling et le transfert des données sont combinés dans une même trame. La *Figure 12* illustre le scénario pour lequel on prévoit également de transmettre chaque trame de données deux fois.

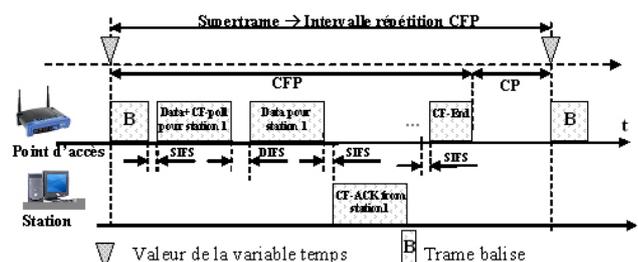

**Figure 12.** *Séquence de transmission avec perte d'une trame : mode PCF*

**Résultats et discussions**

L'application de la compression par la méthode des ondelettes fonctionne et permet de décomposer une image de son choix en sous-bandes.

Les résultats sur la *Figure 14* montrent que nous sommes loin des débits théoriques maximum annoncés par la norme IEEE 802.11. En effet l'ajout d'informations permettant la détection, la synchronisation, la sécurité des données, la gestion, alourdit l'ensemble des données de l'image. Toutefois, ceux-ci permettent de répondre au cahier des charges dans la mesure où la transmission d'une séquence d'images de résolution 512*512 avec une cadence de 10 images/s peut être transmise sans autre traitement : **le débit estimé est d'environ 3,2 Mbit/s**, alors que la bande passante effective pour la norme 802.11b à 11 Mbit/s.

| Image | Résolution | Durée émission d'une image pour 802.11b à 11 Mbps (en ms) | | | Durée émission d'une image pour 802.11g à 54 Mbps (en s) | | |
|---|---|---|---|---|---|---|---|
| | | DCF | DCF RTS/CTS | PCF | DCF | DCF RTS/CTS | PCF |
| IRM | 256*256 | 16,6 | 16,6 | 16,5 | 8,1 | 8,1 | 8,0 |
| IRM | 512*512 | 66,4 | 65,6 | 65,5 | 32,3 | 31,6 | 31,4 |
| Radio. Num. | 2k*2k | 1006,3 | 991,0 | 990,9 | 487,7 | 472,4 | 472,2 |

**Tableau 6.** *Synthèse des temps de transmission d'une image*

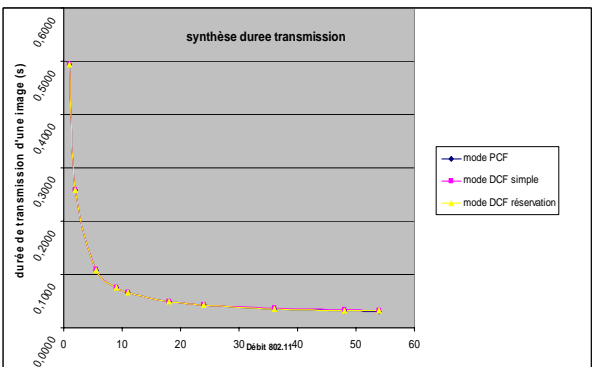

**Figure 13.** *Comparaison de l'ensemble des débits effectifs*

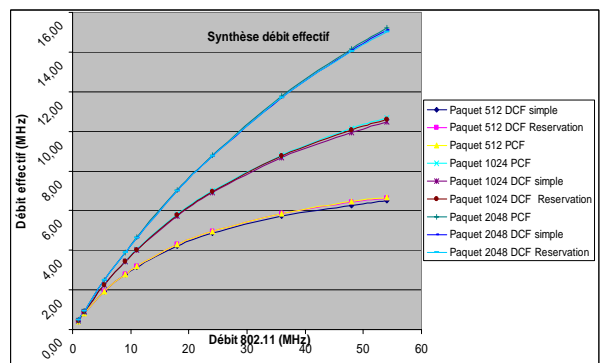

**Figure 14.** *Comparaison de l'ensemble des durées de transmission*

Le *Tableau 6* nous montre que l'écart type des durées nécessaires à la transmission d'une image est très faible et homogène pour chaque valeur de débit du réseau 802.11 et pour trois types d'images.

Pour les images IRM, la norme 802.11b à 11 Mbits/s peut assurer le transfert d'une séquence d'image en respectant la contrainte de 10 images/s. Ce qui n'est toutefois pas possible avec des images de radiologie numérique de 2000*2000 pixels dans la mesure ou la transmission d'une seule image prend une seconde entière.

La durée de la période sans contention à initialiser sur le point d'accès, de manière à permettre la transmission de l'ensemble des données du fichier image dans une seule période, doit être au moins égale à 67 ms. Le point d'accès WRT54G (Linksys) permet le réglage de plusieurs paramètres dont l'intervalle entre deux balises « Beacons ». Nous ne considérons pas dans notre cas de communication autre que la nôtre dans le mode PCF.

On remarque avec les scénarios étudiés que le mode de fonctionnement ne change pas le comportement global de l'application : pour un transfert avec des tailles de paquets et une modulation identiques, la transmission d'une image prend le même temps (à 1 ms près). Le choix du mode de fonctionnement ne dépend donc pas directement du débit attendu mais très certainement d'un critère sur la qualité de la communication dans le cas d'un réseau de plus de deux entités (partage équitable du support, gestion temps réel, limitation des collisions…).

Si l'on voulait transmettre des images de tailles plus importantes, il faudrait alors considérer par exemple le mécanisme détectant les redondances temporelles existant dans la norme MPEG et réduire ainsi la masse des données pour une séquence d'images. Dans le cas d'une séquence d'images de résolution 2000*2000, le débit estimé est de 30,5 Mbps (*Tableau 5*). La transmission d'une seule image se ferait en 1 seconde (*Tableau 6*) pour un débit théorique de 11 Mbps et en 500 ms pour un débit théorique de 54 Mbps, ce qui ne répond plus au cahier des charges que l'on s'est donné car il est impossible de transmettre 10 images par seconde.

La méthode mise au point s'avère adaptée à la compression des images biomédicales. Elle permet d'atteindre, de manière satisfaisante, un taux de compression de 20. Comparés à d'autres codages dont le JPEG, les résultats obtenus par les ondelettes se montrent supérieurs en terme de qualité des images reconstruites.

## 5   Conclusion et perspectives

Ce travail a permis la mise au point d'une méthode efficace de compression d'images médicales. Le caractère irréversible de cette méthode représente aujourd'hui un inconvénient majeur dans la mesure où elle n'offre pas les garanties éthiques et légales de reconstruction exacte de l'image initiale. Cependant les méthodes de compression à perte sont les seules à autoriser des taux de compression élevés. A terme elles constituent une réponse à l'importance grandissante des problèmes liés à la volonté de développer les applications de télémédecine, où disponibilité et rapidité de transfert sont des paramètres critiques. Cette application ne manquera

SETIT2005

pas d'être améliorée dans ses performances de compression, de débit mais aussi de sécurité pour enfin aboutir à un outil de diagnostic à distance réclamé par les praticiens, médecins et bien d'autres.

La norme 802.11 de l'IEEE apparaît comme la mieux adaptée aux réseaux de type WLAN bien que celle-ci soit encore en cours de développement, notamment pour ce qui est de la sécurité des réseaux sans fil haut débit. Pour ce qui est de son fonctionnement, le mode Ad Hoc présente encore des faiblesses puisque aucun déterminisme n'est assuré. En effet, il est impossible de garantir la date de traitement d'une trame.
Bien qu'il offre une liberté moindre pour l'utilisateur et un coût de mise en place supérieur au mode Ad-Hoc, le mode infrastructure est préféré car, contrairement au mode Ad-Hoc, celui-ci garanti un minimum de déterminisme : on est assuré que l'image sera traitée dans le temps imparti et à la cadence de 10 images/seconde dans des conditions moyennes de fonctionnement.

Un temps de super-trame égal à 100 ms permettra le traitement de la séquence d'image comme cela est prévu dans le cahier des charges. Il ne reste alors, qu'à peine la moitié de la bande passante pour le reste des communications en mode DCF.

Quand aux plates-formes de développement, les capacités offertes par les équipements mobiles sont encore loin d'égaler celles d'équipements "standard", ce qui limite fortement les possibilités de développement d'applications réellement performantes.

Une amélioration des résultats obtenus peut se faire pour les technologies de transmission actuelles en travaillant essentiellement au niveau de la couche application, en augmentant le ratio de compression tout en conservant le critère de qualité, mais on peut toutefois espérer des améliorations au niveau de la technologie des réseaux sans fil sur :

- √ L'amélioration de la couverture radio : cela se fait dans les pays qui le permettent en augmentant la puissance. Des travaux réalisés par la COMETA (créée par IBM, AT&T et Intel) devraient permettre d'étendre la couverture sans augmenter la puissance.
- √ L'amélioration du débit : l'IEEE 802.11n prévu pour 2005/2006 permettra des débits théoriques de 108 Mbits/s et même jusqu'à 320 Mbits/s.
- √ Les nouvelles topologies : le mode avec ou sans point d'accès qui permet de réaliser un réseau où les équipements ne peuvent communiquer qu'avec leurs voisins immédiats. Les terminaux effectuent un routage pour acheminer les données de proche en proche.